\documentclass[aps,prb,preprint,superscriptaddress]{revtex4}

\usepackage{epsfig} 

\begin{document}

\title{In-Field  Critical Current of Type-II Superconductors Caused  by
Strain from Nanoscale Columnar Inclusions}

\author{J.P. Rodriguez}
\affiliation{Department of Physics and Astronomy, 
California State University, Los Angeles, California 90032}

\author{P.N. Barnes}
\affiliation{Air Force Research Laboratory, 
Wright-Patterson Air Force Base, Dayton, OH 45433}

\author{C.V. Varanasi}
\affiliation{University of Dayton Research Institute, Dayton, OH 45469}

\date{\today}

\begin{abstract}
The results of a linear elasticity analysis yields that nano-rod inclusions 
aligned along the $c$ axis
of a  thin film
of YBa$_2$Cu$_3$O$_{7-\delta}$,
such as BaZrO$_3$ and BaSnO$_3$,
squeeze that matrix by pure shear.
The 
sensitivity of the superconducting  
critical temperature in that material to the latter
implies that the phase boundary separating the nano-rod inclusion from the superconductor
acts as a collective pinning
center for the vortex lattice that appears in external magnetic field.  
A dominant contribution to the in-field critical current can result.
The elasticity analysis also finds that the growth of 
nano-rod inclusions can be weakly metastable when
the inclusion is softer than the matrix.
\end{abstract}

\maketitle

{\it Introduction.}
The ongoing development of thin films
of superconducting YBa$_2$Cu$_3$O$_{7-\delta}$ (YBCO) for wire technology
has resulted in world-record high critical currents.\cite{review}
In external magnetic field,
the critical current
is considerably enhanced by nano-rod inclusions 
that are aligned in parallel to the crystalline 
$c$ axis.\cite{bzo}$^,$\cite{goyal}$^,$\cite{bso_07}$^,$\cite{mele08}
The enhancement
is strongest at orientations of the 
magnetic field parallel to the $c$ axis.
Understanding the fundamental physics behind this effect remains a challenge.
It is also unknown what drives the growth of nanorods in the first place
in YBCO films.

\begin{figure}
\includegraphics[scale=0.67, angle=-90]{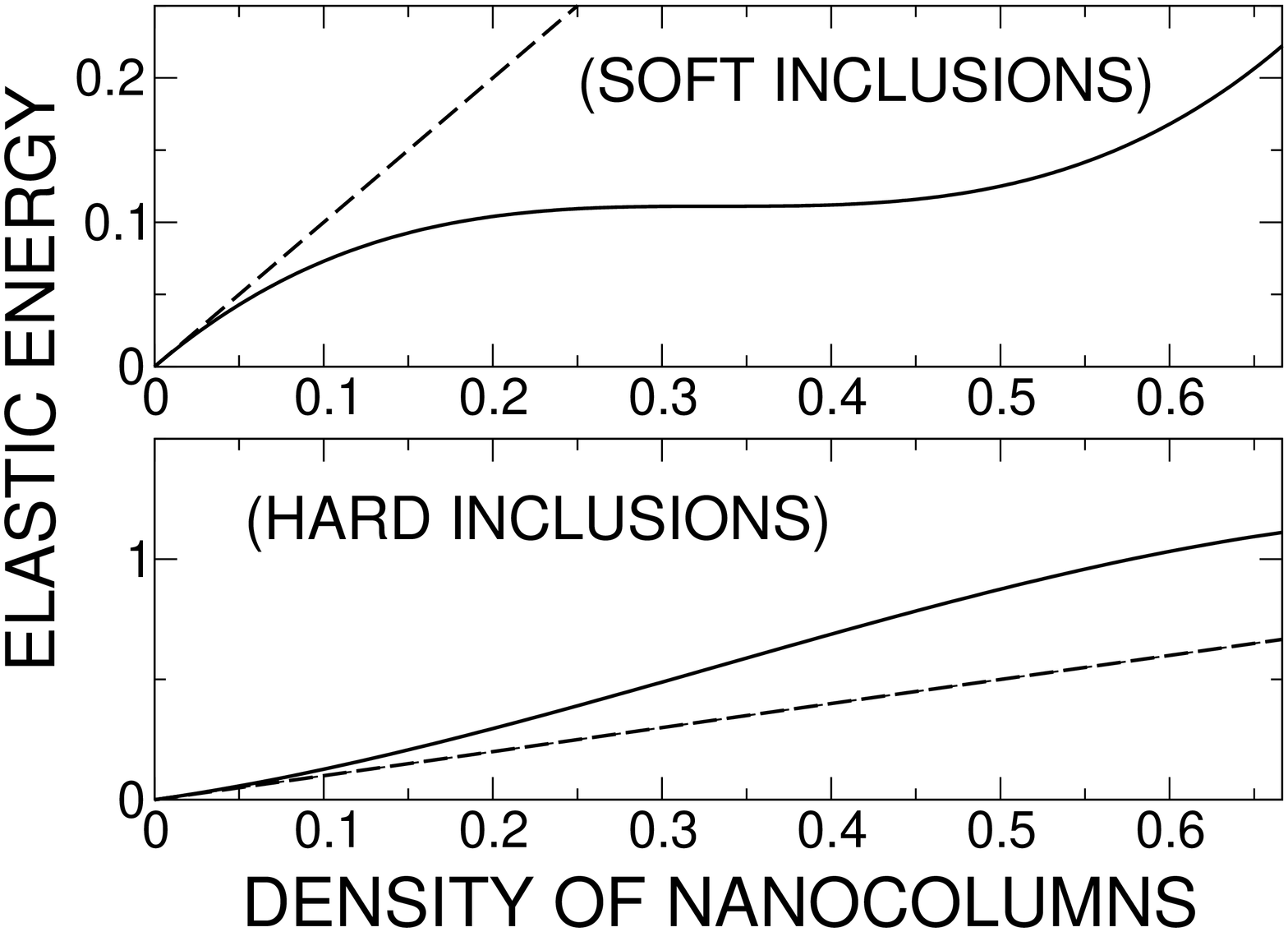}
\caption{Plotted is the total elastic energy density (\ref{energy}),
in units of 
$E_{\rm 2D}^{(1)}/\pi r_1^2$,
versus the density of nano-column inclusions, 
in units of $1/\pi r_1^2$.
The dashed line above corresponds to the elastic energy of isolated nano-column inclusions.
The radii
in Eq. (\ref{energy})
are set to $r_2^2 = \pm 3 r_1^2$ for relatively soft and hard nanocolumns, respectively.}
\label{plot_energy}
\end{figure}

In this paper,
we provide insight into both of these questions by computing the strain field 
due to nano-rod inclusions that thread a YBCO
superconductor along the $c$ axis.  The lattice constant of inclusions
that optimize the critical current
is typically 8\%
larger than that of the YBCO matrix in the $a$-$b$ plane.
Assuming a coherent phase boundary between the inclusion and a given epitaxial layer of YBCO,
a linear elasticity analysis yields that the nanocolumn is compressed axially, while the
YBCO matrix is squeezed by pure shear about the nanocolumn.
The critical temperature
in optimally doped YBCO
is known to couple strongly 
to pure shear in the $a$-$b$ plane.\cite{welp}
In applied magnetic field,
we show theoretically how this experimental fact
results in substantial collective pinning of the vortex lattice
by the phase boundary separating the  nano-column inclusion from the 
YBCO matrix.\cite{jpr-maley}$^,$\cite{jpr07}
Also, the elastic energy shows weak metastability at a high density of nanocolumns
when the nanocolumn is soft compared to the YBCO layer (see Fig. \ref{plot_energy}).
We believe that this drives epitaxial growth of nano-rod inclusions in YBCO films.

{\it Two-Dimensional Elasticity Theory.}
We shall determine first the elastic strain and the elastic energy cost due to 
a  single nano-rod inclusion that threads a film of YBCO along the $c$ axis.
Such nanorods are typically composed either of BaZrO$_3$ (BZO)\cite{bzo}$^,$\cite{goyal} 
or of BaSnO$_3$ (BSO).\cite{bso_07} 
Both are cubic perovskites,
with lattice constants
($a_{\rm in}$)
that exceed that of the $a$-$b$ plane in YBCO,
$a_{\rm out} = 3.86\, {\rm \AA}$,
by 9\% and by 7\% respectively.\cite{bzo}$^,$\cite{goyal}$^,$\cite{bso_07}
Let us temporarily ignore the effect of the lattice mismatch along the $c$ axis
by considering only epitaxial layers that are far from any possible
partial misfit dislocation, and that therefore present
a coherent phase boundary between the inclusion and the YBCO matrix. 
Such partial misfit dislocations are accompanied by stacking faults,\cite{book}  
a topic which will be discussed later in the concluding section.
The assumption of a coherent phase boundary is valid for a nano-rod inclusion of diameter less than the
distance between possible misfit dislocations,\cite{book} 
$a_{\rm Moire} = (a_{\rm out}^{-1} - a_{\rm in}^{-1})^{-1}$.
BZO nanorods
typically have  
a diameter of\cite{goyal} $2{\rm -}3\, {\rm nm}$,
which satisfies the bound $a_{\rm Moire} = 5  \, {\rm nm}$.  
BSO nanocolumns, on the other hand,
typically have 
a diameter of\cite{bso_07}  $7{\rm -}8\, {\rm nm}$.
It exceeds $a_{\rm Moire} = 6  \, {\rm nm}$, although not by much.

Consider then a cylindrical nano-column inclusion that presents a coherent phase boundary with
a given epitaxial layer of the YBCO matrix.
Unit cells match up one-to-one across the phase boundary in such case.
The ideal axial symmetry, assumed here for simplicity,
implies a radial displacement  field,
${\bf u} ({\bf r}) = u(r) {\bf\hat r}$.   We then have the boundary condition
\begin{equation}
u_{\rm out} (r_{\rm out}) - u_{\rm in} (r_{\rm in}) = r_{\rm in} - r_{\rm out}
\label{BC_1}
\end{equation}
between the displacement fields of the nanocolumn (in) and of the YBCO layer (out) at the
phase boundary.  The in-plane lattice mismatch that it represents generates elastic
strain in both the inclusion and in the YBCO matrix.
The elastic energy due to a 2D strain field is given by the integral\cite{book}
\begin{equation}
E_{\rm 2D} = \int^{\prime} d^2 r \Biggl\{ {1\over 2} c_{\parallel} ({\boldmath\nabla}\cdot{\bf u})^2
+ {1\over 2} c_{\perp} 
\Biggl[\Biggl({\partial u_x\over{\partial x}} - {\partial u_y\over{\partial y}}\Biggr)^2
+\Biggl({\partial u_x\over{\partial y}} + {\partial u_y\over{\partial x}}\Biggr)^2\Biggr]\Biggr\}
\label{E_2d}
\end{equation}
over the corresponding area (prime),
which is confined to  $r < r_{\rm in}$ for the nanocolumn
and 
to $r > r_{\rm out}$ for the YBCO matrix.
Here,
$c_{\parallel}$ and $c_{\perp}$ are the 2D bulk compression modulus
and the 2D shear modulus, respectively.
A useful identity for the pure shear component above reads
%
\begin{equation}
\Biggl({\partial u_x\over{\partial x}} - {\partial u_y\over{\partial y}}\Biggr)^2
+\Biggl({\partial u_x\over{\partial y}} + {\partial u_y\over{\partial x}}\Biggr)^2
=
2({\boldmath\nabla}\, {\bf u})^2 - ({\boldmath\nabla}\cdot{\bf u})^2 - ({\boldmath\nabla}\times{\bf u})^2.
\label{identity}
\end{equation}
The strain tensor takes the form 
${\boldmath\nabla} {\bf u} = 
(du/dr) {\bf\hat r}\, {\bf\hat r} + (u/r) {\boldmath\hat\phi}\, {\boldmath\hat\phi}$
in the present axially symmetric case.
It combined with Eq. (\ref{identity}) results in the compact expression for the elastic energy,
$E_{\rm 2D} = \int d^2 r\{{1\over 2} c_{\parallel} [r^{-1} d (r u) /d r]^2
+ {1\over 2} c_{\perp} [r\, d (r^{-1} u) / d r]^2\}$.
Calculus of variations
then yields a  nano-column inclusion squeezed by pure compression 
and a surrounding YBCO matrix squeezed by pure shear:
\begin{equation}
u_{\parallel} (r) = - A_0 r \quad {\rm for} \quad r < r_{\rm in}, 
\quad {\rm and} \quad 
u_{\perp} (r) = + B_0 r_{\rm out}^2 / r \quad {\rm for} \quad r > r_{\rm out},
\label{sol}
\end{equation}
with corresponding  strain tensors
\begin{equation}
{\boldmath\nabla} {\bf u}_{\parallel} = - A_0 {\bf I}
\quad {\rm and} \quad
{\boldmath\nabla} {\bf u}_{\perp} = B_0 (r_{\rm out}/r)^2 
({\boldmath\hat\phi}\, {\boldmath\hat\phi} - {\bf\hat r}\, {\bf\hat r}).
\label{strain}
\end{equation}
%
The total elastic energy  (\ref{E_2d}) generated by the nano-column inclusion is then
$E_{\rm 2D}^{(1)} = 2 c_{\parallel}^{({\rm in})} \pi r_{\rm in}^2 A_0^2
+2 c_{\perp}^{({\rm out})} \pi r_{\rm out}^2 B_0^2$.  
Minimizing it with respect to the constants $A_0$ and $B_0$
while enforcing the boundary condition (\ref{BC_1}) yields 
optimal values
$r_{\rm in} A_0 = (\Delta r) c_{\perp}^{({\rm out})} / (c_{\parallel}^{({\rm in})} + c_{\perp}^{({\rm out})})$ and
$r_{\rm out} B_0 = (\Delta r) c_{\parallel}^{({\rm in})} / (c_{\parallel}^{({\rm in})} + c_{\perp}^{({\rm out})})$. 
Here $\Delta r = r_{\rm in} - r_{\rm out}$. 
These then yield an elastic energy cost
\begin{equation}
E_{\rm 2D}^{(1)} = 2 \pi (\Delta r)^2 (c_{\parallel}^{({\rm in}) -1} + c_{\perp}^{({\rm out}) -1})^{-1}
\label{e1}
\end{equation}
for the nano-column inclusion,
which has an equilibrium radius
$r_{\rm out} + u_{\rm out}(r_{\rm out})$
given by
$r_0 = (c_{\parallel}^{({\rm in})} r_{\rm in} + c_{\perp}^{({\rm out})} r_{\rm out})/
(c_{\parallel}^{({\rm in})} + c_{\perp}^{({\rm out})})$.

Consider next a field of many cylindrical nano-column inclusions
of radius $r_0$ centered at  transverse locations $\{{\bf R}_n\}$.
Suppose again that they all present a coherent phase boundary with 
a given epitaxial layer of the YBCO matrix.
The displacement field is then a linear superposition of those generated
by a single nano-column inclusion (\ref{sol}):
\begin{equation}
{\bf u}_{\rm in} ({\bf r}) = {\bf u}_{\parallel} ({\bf r} - {\bf R}_i)
+ \sum_{j\neq i} {\bf u}_{\perp} 
[(r_{\rm out}/r_{\rm in})({\bf r} - {\bf R}_i) + {\bf R}_i - {\bf R}_j]
\label{sol_in}
\end{equation}
inside the $i^{\rm th}$ nanocolumn, and
\begin{equation}
{\bf u}_{\rm out} ({\bf r}) = \sum_j {\bf u}_{\perp} ({\bf r} - {\bf R}_j)
\label{sol_out}
\end{equation}
inside the YBCO matrix.
The pure shear terms that have been added to the pure compression inside of a nanocolumn (\ref{sol_in})
are required by the boundary condition
(\ref{BC_1}).
Observe now,
by Eq. (\ref{strain}),
that
$\nabla^2{\bf u}_{\parallel} = 0 = \nabla^2{\bf u}_{\perp}$.
Inspection of the elastic energy functional (\ref{E_2d}) combined with
the identity (\ref{identity}) then yields 
that the 
above superpositions
are stationary
because ${\boldmath\nabla}\cdot {\bf u}_{\perp}$,
${\boldmath\nabla}\times {\bf u}_{\parallel}$ and ${\boldmath\nabla}\times {\bf u}_{\perp}$
all vanish.
Indeed,
the elastic energy cost reduces to a sum of surface integrals around the phase boundaries 
of the form
$E_{\rm 2D} = \sum_i E_{\rm 2D}^{(1)}  +
\sum_i \sum_j^{\prime} [e_{i,j,i} ({\rm out}) + e_{i,i,j} ({\rm out})] +
\sum_i \sum_{j,k}^{\prime} [e_{i,j,k}({\rm in}) + e_{i,j,k} ({\rm out})]$,
where the indices $j$ and $k$ refer to the terms 
in the superpositions
(\ref{sol_in}) and (\ref{sol_out}), 
and where the index $i$ refers to the phase  boundary.
The prime notation over the summation symbols indicates that $i\neq j, k$.
Each individual contribution $e_{i,j,k}$
is given by
a surface  integral around the circle 
$S_i$ of radius $r_{\rm out}$ that is centered at ${\bf R}_i$:
$e_{i,j,k} (X) = {\rm sgn} (X)\, c_{\perp}^{(X)} I_{i,j,k}$,
 with
\begin{equation}
I_{i,j,k} = \oint_{S_i}
d {\bf a} \cdot [{\boldmath\nabla} {\bf u}_{\perp} ({\bf r} - {\bf R}_j)]
\cdot {\bf u}_{\perp} ({\bf r} - {\bf R}_k).
\label{surface_int}
\end{equation}
%
Here, ${\rm sgn} ({\rm in}) = + 1$ and ${\rm sgn} ({\rm out}) = - 1$.  
Also, the measure $d{\bf a}$ on the circle $S_i$ points radially outward.  
Substituting in the strain fields (\ref{strain}) above yields ultimately
that $e_{i,j,i} = 0 = e_{i,i,j}$, and 
that
\begin{equation}
e_{i,j,k} (X) = 
{\rm sgn} (X) c_{\perp}^{(X)}
 (2\pi) B_0^2 r_{\rm out}^6
{\rm Re}\, [R_{i,j} R_{i,k} e^{i\phi_{j,k} (i)} - r_{\rm out}^2]^{-2}
\label{3-body}
\end{equation}
for $i\neq j,k$.  
(See Appendix.)
Here, ${\bf R}_{i,j} = {\bf R}_i - {\bf R}_j$,
and $\phi_{j,k} (i)$ denotes the angle between the vectors ${\bf R}_{i,j}$ and ${\bf R}_{i,k}$.
The 2D elastic energy then is composed of a sum 
of 1-body ,
2-body ($j = k$) and 3-body terms ($j \neq k$),
$E_{\rm 2D} = \sum_i E_{\rm 2D}^{(1)}  + \sum_i \sum_{j,k}^{\prime} V_{i,j,k}$,
with the  interaction energy given by
\begin{equation}
V_{i,j,k}  =
- (2\pi) [c_{\perp}^{({\rm out})} - c_{\perp}^{ ({\rm in})}] B_0^2 r_{\rm out}^6
{\rm Re}\, [R_{i,j} R_{i,k} e^{i\phi_{j,k} (i)} - r_{\rm out}^2]^{-2}.
\label{interaction}
\end{equation}
Notice that  $V_{i,j,k}$
changes sign as a function of the relative rigidity between 
the nano-column inclusion and the YBCO matrix.

The elastic energy will now be obtained by computing subsequent self-energy corrections to 
the 2-body interaction
and to the 1-body line tension.
Let's first fix the  coordinate for the phase boundary above, ${\bf R}_i$, 
as well as one of the  nanocolumn coordinates above, ${\bf R}_j$.  
Observe that the 3-body interaction
(\ref{interaction}) has zero angle average about the center ${\bf R}_i$
over the remaining nanocolumn coordinate ${\bf R}_k$.
This is due simply to the fact that the contour integral
$\oint dz z^{-1} (z - w)^{-2}$ around the unit circle,
$z = {\rm exp}[i\phi_{j,k} (i)]$,
vanishes for complex $w$ inside of that circle.
Let's assume that each nanocolumn has a hard core of radius $r_1 \sim  r_0$.
At $R_{i,j} \gg r_1$,
we then  obtain the estimate
$\sum_{k}^{\prime} V_{i,j,k} = - \pi r_1^2\, n_{\phi} V_{i,j,j}$
for the correction to the 2-body interaction
on average over the bulk of the system.
Here, $n_{\phi}$ denotes the density of nanocolumns.
The renormalized 2-body interaction that results is then
$V_{i,j}^{(2)} = (1 - \pi r_1^2\, n_{\phi}) V_{i,j,j}$.
Next, assume an effective hard-core of radius $r_2^{\prime} \sim r_1$ 
for the nanocolumn at the coordinate ${\bf R}_j$ that remains.
We thereby obtain the estimate
$\sum_{j,k}^{\prime} V_{i,j,k} = 
n_{\phi} \int^{\prime} d^2 R_{i,j} V_{i,j}^{(2)}
= - \pi r_2^{2} n_{\phi} (1 - \pi r_1^2\, n_{\phi}) E_{\rm 2D}^{(1)}$
for the net self-energy correction to the elastic energy of an isolated nano-column inclusion,
with
%
\begin{equation}
r_2^{2} = [(1 - c_{\perp}^{({\rm in})}/c_{\perp}^{({\rm out})}) /
(1 + c_{\perp}^{({\rm out})} / c_{\parallel}^{({\rm in})})]
\cdot
[r_{\rm out}^4 /
(r_2^{\prime 2} - r_{\rm out}^2)].
\label{r_2}
\end{equation}
This yields a total elastic energy density
\begin{equation}
E_{\rm 2D}/A = [1 -\pi r_2^{2}\, n_{\phi} (1 - \pi r_1^2\, n_{\phi})]
 n_{\phi} E_{\rm 2D}^{(1)}
\label{energy}
\end{equation}
as a function of the density of nanocolumns. 
The above third-order polynomial is depicted by Fig. \ref{plot_energy}.
It notably predicts weakly metastable
epitaxial growth for
relatively soft nanorods within the YBCO matrix, such that
$c_{\perp}^{({\rm in})} < c_{\perp}^{({\rm out})}$.
This occurs at a density
$n_{\phi} = (1 + [1 - (3 r_1^2/r_2^2)]^{1/2})/3\pi r_1^2$
of nano-rod inclusions,
at large effective crossections
$\pi r_2^{2} > 3 \pi r_1^2$.
The equilibrium density of nano-rod inclusions therefore {\it cannot} be dilute.
In particular, 
$3 \pi r_1^2 n_{\phi}$
must lie  somewhere between $1$ and $2$.
Inspection of Eq. (\ref{r_2}) indicates that
the former condition requires some degree of agglomeration among the nano-column inclusions:
$r_{\rm out} < r_2^{\prime}  < 2 r_{\rm out}$.
This may, however, be an artifact of the previous estimate for the 2-body self-energy correction,
which is not accurate at $R_{i,j}\sim r_1$.
Last, the elastic energy cost per unit volume (\ref{energy}) at meta-stable equilibrium
is 
$E_{\rm 2D}^{(1)}/9\pi r_1^2 = (2/9)(\Delta r / r_1)^2
(c_{\parallel}^{({\rm in}) -1} + c_{\perp}^{({\rm out}) -1})^{-1}$
in the marginally stable limit
at $r_2^2 = 3 r_1^2$ (see Fig. \ref{plot_energy}).
The strong dependence that it shows on the bulk compression modulus of the inclusion
affects growth dynamics. This  could be the root cause for the difference in length
between BZO nanorods and BSO nanocolumns in YBCO.\cite{mele08}

{\it Critical Current by Two-Dimensional Collective Pinning.}
We shall now determine the critical current of a thin film of superconducting YBCO
threaded by nano-rod inclusions along the crystalline $c$ axis
and  subject to external  magnetic field aligned along the same axis.
Recall that the critical temperature in an optimally doped YBCO superconductor is primarily sensitive to 
{\it shear} strain in the $a$-$b$ plane.\cite{welp}  
That fact coupled with the shear strain
generated by a nano-column inclusion
(\ref{strain})
results in a potential-energy landscape
for vortex lines that can collectively pin the 
vortex lattice.
In particular,
the contribution of the vortex core to the vortex line tension
is approximated by the fundamental energy scale per unit length 
$\varepsilon_0 = (\Phi_0/4\pi \lambda_L)^2$,
where $\lambda_L$ denotes the London penetration depth.
The temperature dependence shown by the  vortex line tension is therefore  approximated by
$\varepsilon_0 (T) = \varepsilon_0 (0) [1 - (T/T_{c0})]$
near the mean-field critical temperature $T_{\rm c0}$.
The potential-energy landscape experienced by a  vortex line then has a contribution
$\delta\varepsilon_1 ({\bf r}) = \sum_{\alpha}\sum_{\beta} 
(\partial\varepsilon_0 / \partial{T_{c}})
(\partial{T_{c}} / \partial\epsilon_{\alpha , \beta })
\epsilon_{\alpha , \beta } ({\bf r})$,
where $T_c$ is the true critical temperature, and where
$\epsilon_{\alpha , \beta }$
is the symmetric strain tensor (\ref{strain}).
It results in a $d$-wave  potential-energy landscape
about the nanocolumn for a vortex core,
\begin{equation}
\delta\varepsilon_1 ({\bf r}) = \varepsilon_{\rm p} (r_{\rm out}/r)^2 {\rm cos}\, 2\phi \ ,
\label{potential}
\end{equation}
with
$\varepsilon_{\rm p} = \varepsilon_0 (0) (T/T_{c0}) 
T_c^{-1}[(\partial{T_{c}} / \partial\epsilon_{bb}) - (\partial{T_{c}} / \partial\epsilon_{aa})] B_0$.
Here  the ratio between 
$T_{c0}$ and $T_{c}$ 
is  assumed to be constant.
A rigid vortex line therefore  experiences a  force field 
\begin{equation}
{\bf f}_1 ({\bf r}) = f_{\rm p} (r_{\rm out} / r)^3 
({\bf\hat r}\, {\rm cos}\, 2\phi + {\bf\hat\phi}\, {\rm sin}\, 2\phi)
\label{force}
\end{equation}
due to the strain generated by a single nano-column inclusion,
where $f_{\rm p} = 2\varepsilon_{\rm p} /r_{\rm out}$ is the 
maximum
force per unit length.

\begin{figure}
\includegraphics[scale=0.8, angle=-90]{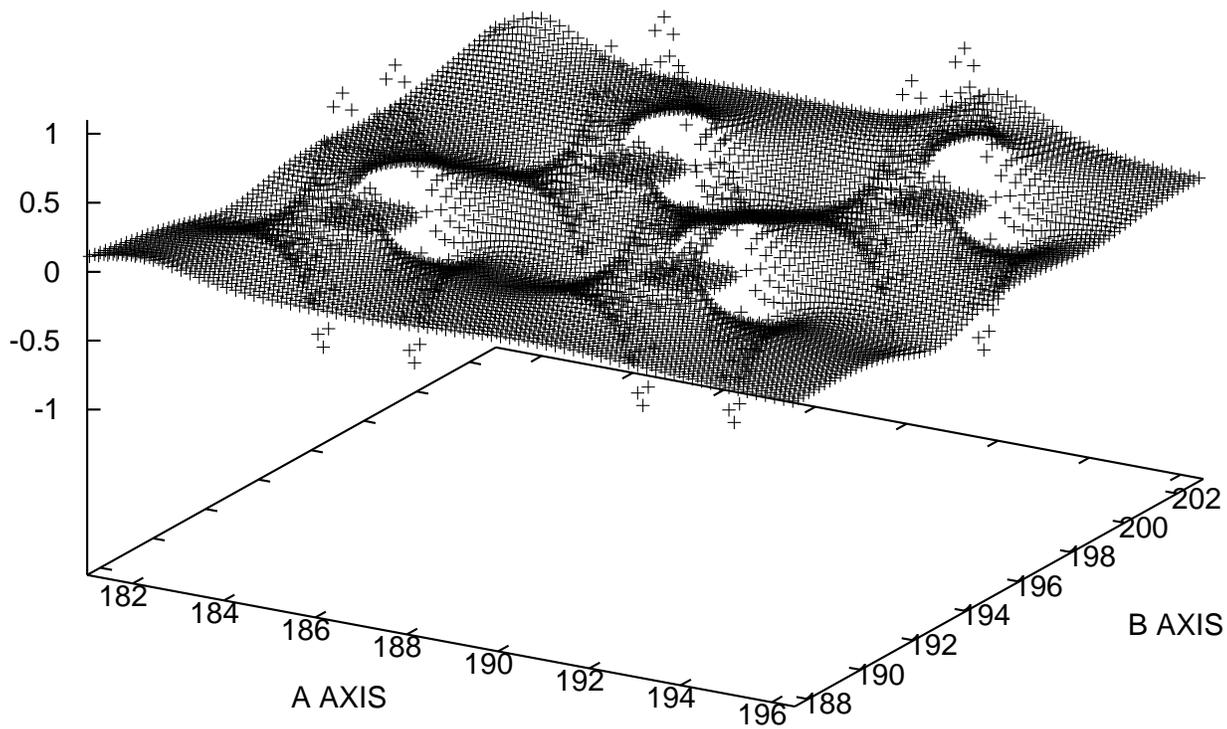}
\caption{Displayed is a potential-energy landscape in units of
$|\varepsilon_{\rm p}|$
and of the coherence length
for a single vortex line 
that results from a superposition
of 2744 $d$-wave collective-pinning centers [Eq. (\ref{potential})]
arranged in a ``liquid'' fashion. (See ref. 8.)}
\label{plot_landscape}
\end{figure}

The above pinning/anti-pinning force (\ref{force}) 
is long range.  
The presence of an extended field of nanocolumns can 
cut the range off, however. 
(See Fig. \ref{plot_landscape}.)
Such forces add within the present elastic approximation (\ref{sol_out}):
${\bf f} ({\bf r}) = \sum_i {\bf f}_1 ({\bf r} - {\bf R}_i)$.
The $d$-wave nature of each isolated force field (\ref{force}) implies a null net force on average.
A characteristic fluctuation of the force  over the YBCO matrix remains:
$\overline{f^2} = n_{\phi} \int^{\prime} d^2 r |{\bf f}_1 ({\bf r})|^2 = {1\over 2} (\pi r_{\rm out}^2 n_{\phi}) f_{\rm p}^2$,
where integration (prime) is restricted to the YBCO matrix.
Matching
$\overline{f^2}$
with
$|{\bf f}_1 ({\bf r})|^2$
yields an effective range for each pinning/anti-pinning center
$r_{\rm p} =(2/\pi n_{\phi})^{1/6} r_{\rm out}^{2/3}.$  

The $d$-wave potential (\ref{potential})
that acts on rigid vortex lines in the vicinity of
the phase boundaries between the nano-column inclusions and  the YBCO matrix 
has zero angle average.
It therefore cannot pin down a  vortex line in isolation.
Previous work by one of the authors and Maley\cite{jpr-maley} implies that 
many of them
collectively pin the Abrikosov vortex lattice, however.
A hexatic Bose glass state can exist at low temperature.\cite{jpr07}
It is a  vortex lattice threaded by
isolated lines of edge dislocations
in parallel to the 
relatively weak correlated pinning/anti-pinning centers.
Plastic creep of the vortex lattice
associated with glide by such edge dislocations
limits the critical current,$\cite{jpr-maley}$
which is given by
$j_c  B / c  \sim n_{\rm p} f^2_{\rm p} / c_{66} b$.
Here $n_{\rm p}$
denotes the density of vortex lines 
pinned by the nanocolumns,
$c_{66}=(\Phi_0 / 8 \pi \lambda_L)^2 n_B$ is
the elastic shear modulus of the pristine vortex lattice
at a density $n_B$ of vortex lines,\cite{brandt77}
and $b$ denotes the magnitude of the  Burgers vector 
associated with the edge dislocations that thread the vortex lattice.
The $d$-wave nature of the 
pinning/anti-pinning center
 (\ref{potential})
also implies that its occupation is purely random.
The density of vortex lines that they collectively pin
is then equal to 
$n_{\rm p} = (\sigma_{\rm p} n_B) n_{\phi}$, where
$\sigma_{\rm p} = \pi (r_{\rm p}^2 - r_{\rm out}^2)$
is the  effective crossectional area of a pinning/anti-pinning center
(see Fig. \ref{plot_landscape}).  
The critical current density therefore
obeys a {\it pure} inverse-square-root power law with magnetic field, $j_c\propto B^{-1/2}$.
Taking values 
of $\partial{T_{c}} / \partial\epsilon_{aa} = 230\,{\rm K}$ and
$\partial{T_{c}} / \partial\epsilon_{bb} = - 220\,{\rm K}$ 
for the strain derivatives of $T_c$ in optimally-doped YBCO\cite{welp}
can result in a pinning efficiency, 
$|f_p| \xi /\varepsilon_0$,
of 93\%  at liquid nitrogen temperature!


{\it Discussion and Conclusions.}
We have found that the growth of nano-rod inclusions in YBCO films is
very likely driven by weak metastability shown by the elastic energy of epitaxial layers.  
We also have pointed out how the 
sensitivity of the critical temperature
in optimally-doped YBCO to pure shear strain inside of the $a$-$b$ plane\cite{welp} results
in an effective collective pinning center for the Abrikosov  vortex lattice
at the phase boundary between the nano-rod inclusion  and the YBCO matrix.

The lattice mismatch along the $c$ axis between the nano-rod inclusion and YBCO has so far been 
neglected, however.
YBCO has a unit cell that can be divided into a stack of three 
cubes along the $c$ axis,
each with a lattice constant $c_{\rm out} / 3 = 3.9\,{\rm \AA}$.
The strain that results at the phase boundary with a BZO nanorod
or with a BSO nanocolumn,
both of which are cubic with lattice constants 
$a_{\rm in} = 4.2\, {\rm \AA}$ and $4.1\, {\rm \AA}$, respectively,
can be relieved by introducing partial misfit dislocations
accompanied by stacking faults in the YBCO matrix.\cite{book}
The predicted spacing between such stacking faults,
$c_{\rm Moire} = [(3/c_{\rm out}) - a_{\rm in}^{-1}]^{-1}$, 
is then equal to $5\,{\rm nm}$ for BZO nanorods
and to $8\, {\rm nm}$ for BSO nanocolumns (cf. ref. 5).
Since their effect on the previous elasticity analysis
can be accounted for by renormalized elastic moduli for the YBCO matrix, 
we believe that that our conclusions remain unchanged in their presence.

The authors thank George Levin for discussions.
This work was supported in part by the US Air Force
Office of Scientific Research under grant no. FA9550-06-1-0479.

\appendix*
\section {Surface Integrals}
Equation  (\ref{surface_int}) gives the surface integral 
that determines the 
3-body elastic interaction
among nano-column inclusions.
Integration by parts combined with $\nabla^2 {\bf u}_{\perp} = 0$
yields that it is symmetric with respect to the latter:
$I_{i,j,k} = I_{i,k,j}$.
In the case that
$i = j$, 
it reduces to the angular integral
\begin{equation}
I_{i,i,k} =  {1\over 2} r_{\rm out}^2 B_0^2
\int_0^{2\pi} d \phi \Biggl(
{R_{i,k}^2  - r_{\rm out}^2\over{r_{i,k}^2}} - 1
\Biggr),
\label{i=j}
\end{equation}
where
$r_{i,k}^2 = r_{\rm out}^2 + R_{i,k}^2 + 2\, r_{\rm out} R_{i,k}\, {\rm cos} (\phi - \phi_k)$.
Here, $\phi_k$ denotes the orientation of the vector ${\bf R}_{i,k} = {\bf R}_i - {\bf R}_k$.
After making the change of variables $z = e^{i\phi}$,
application of Cauchy's theorem
yields that the integral vanishes: $I_{i,i,k} = 0$.
In the case that
$i\neq j$ and $i\neq  k$, the surface integral (\ref{surface_int}) reduces to
\begin{equation}
I_{i,j,k} =  {1\over 4} r_{\rm out}^4 B_0^2
\int_0^{2\pi} d \phi \Biggl[
{(R_{i,j}^2 - r_{\rm out}^2)\over{r_{i,j}^4}} 
- {1\over{r_{i,k}^2}}  
 + {R_{j,k}^2\over{r_{i,j}^2 r_{i,k}^2}}
 - {(R_{i,j}^2 - r_{\rm out}^2) R_{j,k}^2\over{r_{i,j}^4 r_{i,k}^2}}
+ (j\leftrightarrow k)\Biggr].
\label{ijk}
\end{equation}
Repeating the previous steps results in a closed-form expression
with a large number of terms.
Symbolic manipulation programs
then help reduce these to the 
result (\ref{3-body}).

\end{document}